\documentclass[useAMS,usenatbib]{mn2e}

\usepackage{graphicx}

\def\jh{\mbox{$\rm (J-H)$}}

\def\jk{\mbox{$\rm (J-K_s)$}}
\def\mMJ{\mbox{$\rm (m-M)_J$}}

\def\ebv{\mbox{$\rm E(B-V)$}}
\def\ejh{\mbox{$\rm E(J-H)$}}
\def\rc{\mbox{$\rm R_{C}$}}
\def\rl{\mbox{$\rm R_{lim}$}}
\def\rt{\mbox{$\rm R_{t}$}}
\def\ms{\mbox{$\rm M_\odot$}}
\def\ds{\mbox{$\rm d_\odot$}}
\def\mObs{\mbox{$\rm M_{obs}$}}
\def\mTot{\mbox{$\rm M_{tot}$}}

\def\Rgc{\mbox{$\rm R_\odot$}}
\def\dgc{\mbox{$\rm R_{GC}$}}
\def\rx{\mbox{$\rm R_{ext}$}}

\def\jj{\mbox{$\rm J$}}
\def\hh{\mbox{$\rm H$}}
\def\ks{\mbox{$\rm K_s$}}
\def\aV{\mbox{$\rm A_V$}}
\def\ns{\mbox{$\rm N_{1\sigma}$}}

\title[The nature of the GC candidates FSR\,1603 and FSR\,1755]{A populous intermediate-age
open cluster and evidence of an embedded cluster among the FSR globular cluster candidates}

\author[Bica \& Bonatto]{E. Bica$^1$ and C. Bonatto$^1$\\
$^1$ Departamento de Astronomia, Universidade Federal do Rio Grande do Sul\\
Av. Bento Gon\c{c}alves 9500, Porto Alegre 91501-970, RS, Brazil; charles@if.ufrgs.br;
bica@if.ufrgs.br}

\begin{document}


\maketitle


\begin{abstract}
We study the nature of the globular cluster (GC) candidates FSR\,1603 and FSR\,1755 selected from the 
catalogue of \citet{FSRcat}. Their properties are investigated with 2MASS field-star decontaminated 
photometry, which is used to build colour-magnitude diagrams (CMDs), and stellar radial density profiles 
(RDPs). FSR\,1603 has the open cluster (OC) Ruprecht\,101 as optical counterpart, and we show it to be a 
massive intermediate age cluster (IAC). Relevant parameters of FSR\,1603 are the age $\approx1$\,Gyr, 
distance from the Sun $\ds\approx2.7$\,kpc, Galactocentric distance $\dgc\approx6.4$\,kpc, core radius
$\rc\approx1.1$\,pc, mass function slope $\chi\approx1.8$, observed stellar mass (for stars with mass in
the range $\rm 1.27\,\ms\leq m\leq2.03\,\ms$) $\mObs\approx500\,\ms$, and a total (extrapolated to 
$\rm m=0.08\,\ms$) stellar mass $\mTot\approx2300\,\ms$. FSR\,1755, on the other hand, is not a 
populous cluster. It may be a sparse young cluster embedded in the H\,II region Sh2-3, subject to an 
absorption $\aV\approx4.1$, located at $\ds\approx1.3$\,kpc. Important field-star contamination, 
spatially-variable heavy dust obscuration, even in \ks, and gas emission characterise its field. A 
nearly vertical, sparse blue stellar sequence shows up in the CMDs. 
\end{abstract}


\section{Introduction}
\label{intro}

A catalogue containing 1021 star cluster candidates for Galactic latitudes $|b|<20^\circ$ and all
longitudes was recently published by \citet{FSRcat}. Their targets (hereafter FSR objects) were
detected by means of an automated algorithm that basically identified regions with stellar
overdensities, applied to the 2MASS\footnote{The Two Micron All Sky Survey, available at
{\em www.ipac.caltech.edu/2mass/releases/allsky/ }} database. The overdensities were
classified according to a quality flag, '0' and '1' representing the most probable star clusters,
while the '5' and '6' flags may be related to field fluctuations.

By means of a combination of three parameters, the number of cluster stars corrected to a common
magnitude limit, the core radius and the central star density, \citet{FSRcat} could discriminate
known GCs from OCs. When applied to the catalogue, this criterion led them to classify
1012 targets as OC candidates and 9 as GC candidates that should be explored in followup studies.
Among the latter are FSR\,1603 and FSR\,1755.

Several works have already explored the FSR catalogue with varying depths and different approaches,
with results that reflect the importance of such catalogue. The recently discovered GCs FSR\,1735
(\citealt{FSR1735}) and FSR\,1767 (\citealt{FSR1767}), and the probable GCs FSR\,584 (\citealt{FSR584})
and FSR\,190 (\citealt{FSR190}), are clear examples of the fundamental r\^ole played by the FSR catalogue
to improve the statistics of very old star clusters. Indeed, FSR\,1735 and FSR\,1767 are the most recent
additions to the Galactic GC population, a number that presently amounts to $\sim160$ members (e.g.
\citealt{Pap11GCs}).

Besides the obvious importance of finding new star clusters of any age throughout the Milky Way, the 
identification of the IACs FSR\,31, FSR\,89 and FSR\,1744, projected towards
the centre of the Galaxy (\citealt{OldOCs}), has provided as well important information on cluster disruption
processes and survival rates in those regions. In this context, derivation of astrophysical parameters
of new star clusters can be used in studies related to the star formation and evolution processes, dynamics
of N-body systems, cluster disruption time scales, the geometry of the Galaxy, among others.

Central to the problem of identifying the nature of catalogue stellar overdensities is the availability
of a field-star decontamination algorithm to disentangle physical CMD sequences from field fluctuations.
Low-noise stellar radial-density profiles (RDPs) spanning a wide radial range are essential as well, 
especially for objects projected against
dense stellar fields (e.g. \citealt{BB07}). For instance, based on such premises and working with the
decontamination algorithm described in \cite{BB07} applied to 2MASS photometry, \citet{OldOCs} found that
the GC candidate FSR\,89 is rather a well-defined IAC.

Most of the FSR catalogue remains to be further explored. As another step in the direction of
classifying FSR candidates --- taking into account CMD and RDP properties --- we show in the
present work that the candidates FSR\,1603 and FSR\,1755 are not GCs. Instead, FSR\,1603 is a
massive IAC projected $\sim60^\circ$ from the centre of the Galaxy, and FSR\,1755 is an embedded cluster
$\sim12^\circ$ from the centre.

This paper is structured as follows. In Sect.~\ref{targets} we present fundamental data and optical
and near-IR images of FSR\,1603 and FSR\,1755. In Sect.~\ref{PhotPar} we present the 2MASS photometry
and discuss the methods employed in the CMD analyses, especially the field-star decontamination and the
stellar radial density profiles. In Sect.~\ref{Disc} we apply the tools to both candidates and discuss
their nature. Concluding remarks are given in Sect.~\ref{Conclu}.

\begin{figure*}
\begin{minipage}[b]{0.50\linewidth}
\includegraphics[width=\textwidth]{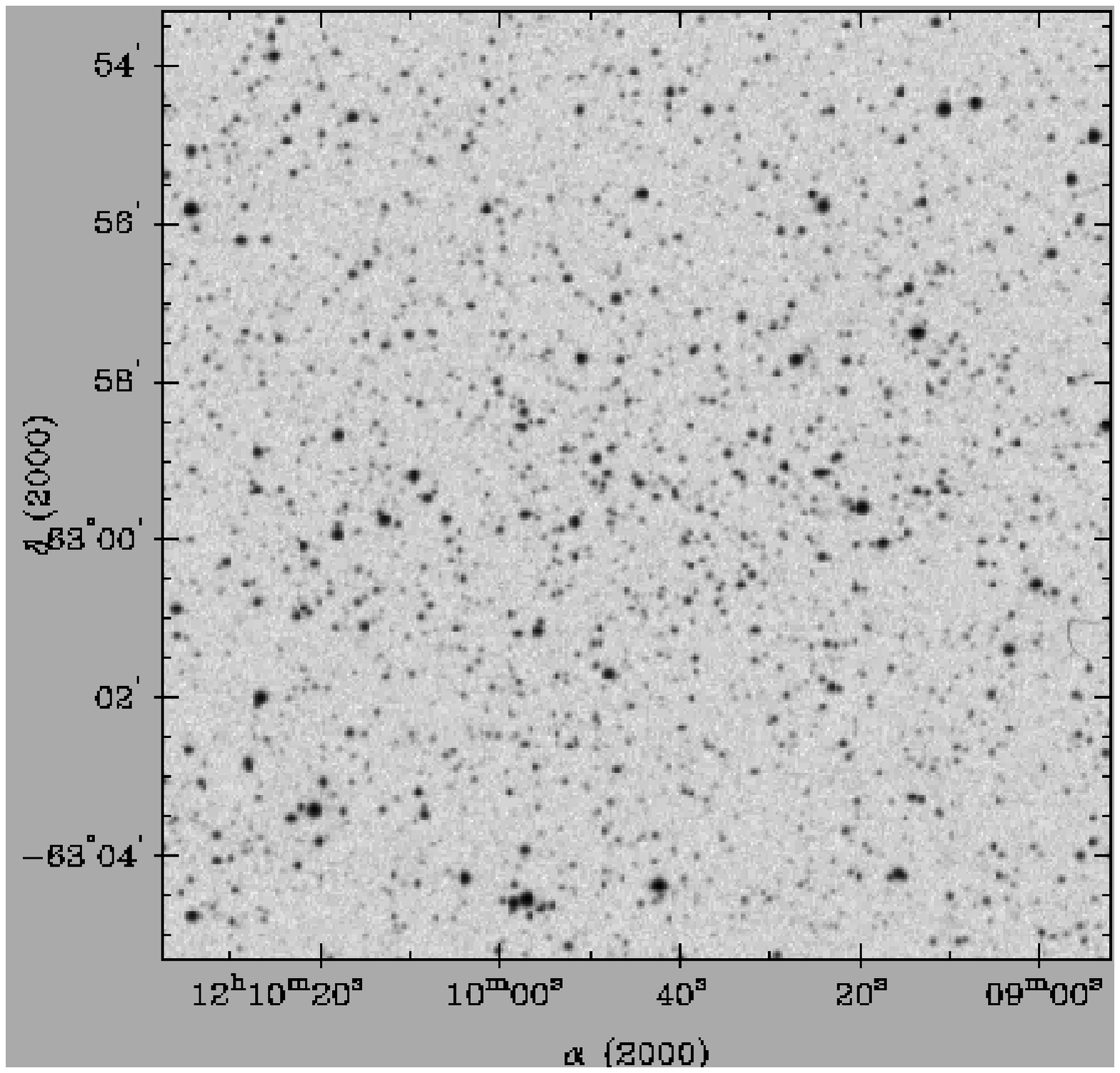}
\end{minipage}\hfill
\begin{minipage}[b]{0.50\linewidth}
\includegraphics[width=\textwidth]{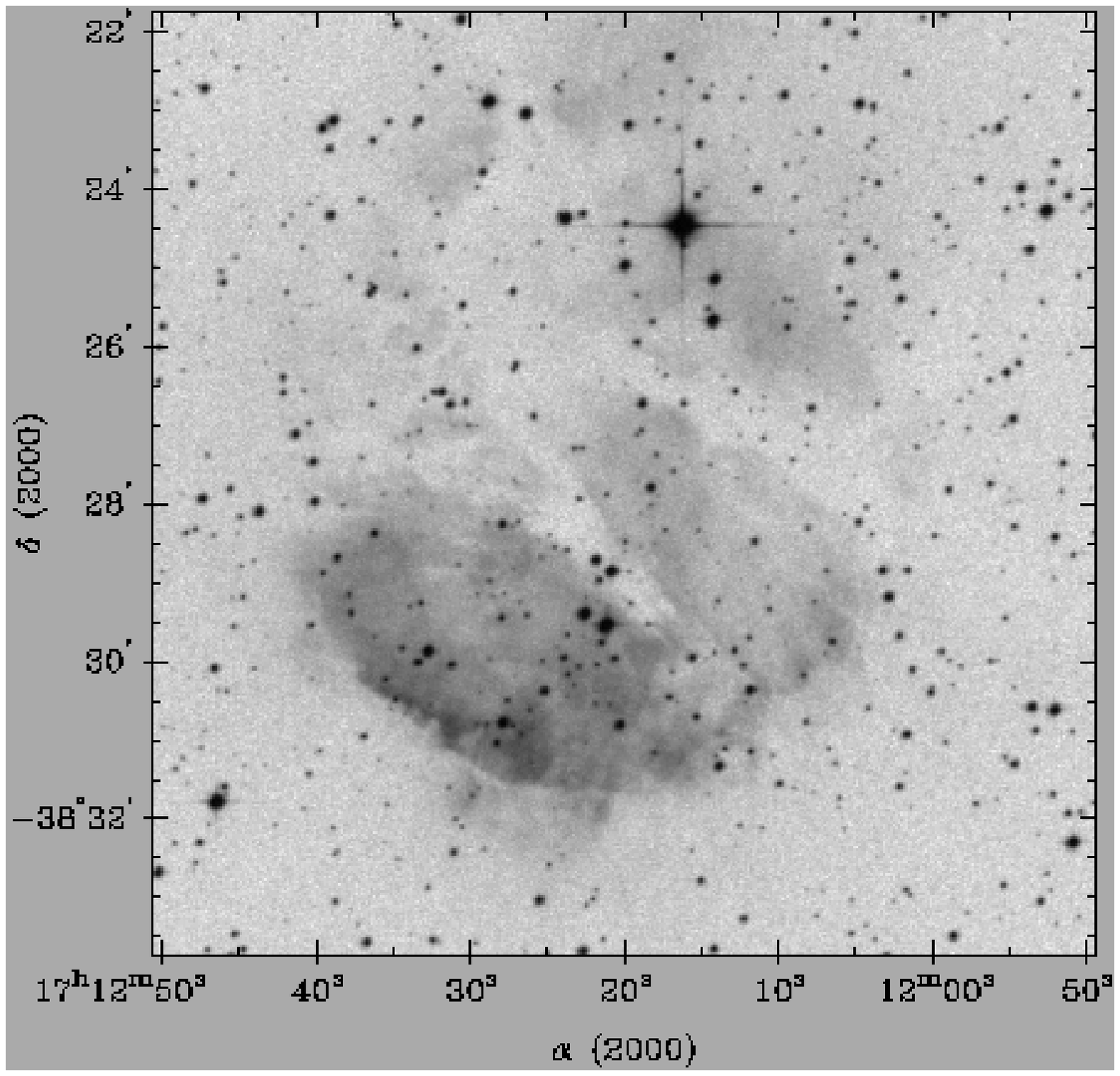}
\end{minipage}\hfill
\caption[]{Left panel: $12\arcmin\times12\arcmin$ DSS B image of FSR\,1603. Right panel:
$12\arcmin\times12\arcmin$ XDSS R image of FSR\,1755, where gas emission and strong dust 
absorption are conspicuous. Images centred on the FSR coordinates (cols.~2 and 3 of 
Table~\ref{tab1}).}
\label{fig1}
\end{figure*}

\section{The GC candidates FSR\,1603 and FSR\,1755}
\label{targets}

Table~\ref{tab1} provides original information on FSR\,1603 and FSR\,1755, where we also include the
core and tidal radii measured by \citet{FSRcat} in the 2MASS \hh\ images by means of a \citet{King1962}
profile fit. Both candidates have been classified as probable star clusters by \citet{FSRcat}; the
quality flags are also given.  The different quality flags of FSR\,1603 and FSR\,1755 can be accounted 
for by the significantly different numbers of prospective cluster members computed by \citet{FSRcat},
which we include in the last column of Table~\ref{tab1}.

FSR\,1603 has the OC Ruprecht\,101 as optical counterpart, for which the coordinates are the only
data provided by the WEBDA\footnote{\em obswww.unige.ch/webda - \citet{Merm1996}} database. The
relatively low contrast of FSR\,1603 with respect to the background can be seen in the
$12\arcmin\times12\arcmin$ DSS\footnote{Extracted from the Canadian Astronomy Data Centre (CADC),
at \em http://cadcwww.dao.nrc.ca/} B image (Fig.~\ref{fig1}, left panel). Nevertheless, a few bright
stars appear to stand out around the image centre over a rather uniform stellar background.

The field of FSR\,1755, on the other hand, presents a complex structure that includes important 
contamination by bulge stars, strong absorption by dust lanes and $\rm H_\alpha$ emission, clearly
seen in the $12\arcmin\times12\arcmin$ XDSS R image (Fig.~\ref{fig1}, left panel). These features
show up as well in the near IR, as can be seen in the 2MASS \ks\ images (Fig.~\ref{fig2}), in two
different spatial scales. Notice the presence of a bright O star about 2\,arcmin to the south of
the central coordinates of FSR\,1755 (Sect.~\ref{FSR1755}).

\begin{table*}
\caption[]{Data from \citet{FSRcat}}
\label{tab1}
\renewcommand{\tabcolsep}{3.0mm}
\renewcommand{\arraystretch}{1.25}
\begin{tabular}{ccccccccccc}
\hline\hline
Object&&$\alpha(2000)$&$\delta(2000)$&$\ell$&$b$&\rc&\rt&Classification&Quality flag&$N_c\ (\hh=15)$\\
&&(hms)&($^\circ\,\arcmin\,\arcsec$)&($^\circ$)&($^\circ$)&(\arcmin)&(\arcmin)&&&(stars)\\
(1)&&(2)&(3)&(4)&(5)&(6)&(7)&(8)&(9)&(10)\\
\hline
FSR\,1603&&12:09:45&$-62$:59:49&298.22&$-0.51$&1.1&29.6&Probable&0&1842 \\
FSR\,1755&&17:12:20&$-$38:27:44&348.25&$+0.48$&3.2& 6.4&Probable&2&90 \\
\hline
\end{tabular}
\begin{list}{Table Notes.}
\item Cols.~2-3: Central coordinates. Cols.~4-5: Corresponding Galactic coordinates. Cols.~6 and
7: Core and tidal radii derived from King fits to the 2MASS \hh\ images. Col.~8: Both candidates
have been considered as probable star clusters. Col.~9: FSR quality flag. Col.~10: Number of 
prospective cluster members, corrected to $\hh=15$.
\end{list}
\end{table*}


\section{Photometry and analytical tools}
\label{PhotPar}

In this section we briefly describe the photometry and outline the methods we apply in the CMD
analyses.

\subsection{2MASS photometry}
\label{2mass}

In both cases, \jj, \hh\ and \ks\ 2MASS photometry was obtained in a relatively wide circular field
of extraction radius $\rx=50\arcmin$ centred on the coordinates provided by \citet{FSRcat} (cols.~2
and 3 of Table~\ref{tab1}) using VizieR\footnote{\em vizier.u-strasbg.fr/viz-bin/VizieR?-source=II/246}.
Wide extraction areas can provide the required statistics, in terms of magnitude and colours, for a
consistent field star decontamination (Sect.~\ref{CMDs}). They are essential as well to produce stellar
RDPs with a high contrast with respect to the background (Sect.~\ref{RDPs}). 
In the case of FSR\,1603, the RDP resulting
from the original FSR coordinates presented a dip at the centre. Thus, we searched for new coordinates 
to maximise the star-counts in the innermost RDP bin. The optimised central coordinates are given in 
cols.~2 and 3 of Table~\ref{tab2}. The offset with respect to the original coordinates is relatively
small, $\Delta\alpha=-6.0\arcsec$ and $\Delta\delta=-32.2\arcsec$.

As a photometric quality constraint, the 2MASS extractions were restricted to stars with errors in \jj,
\hh\ and \ks\ smaller than 0.25\,mag. A typical distribution of uncertainties as a function of magnitude,
for objects projected towards the central parts of the Galaxy, can be found in \citet{BB07}. About $75\% -
85\%$ of the stars have errors smaller than 0.06\,mag.

\subsection{Colour-magnitude diagrams and field-star decontamination}
\label{CMDs}

The CMD morphology is an essential tool to unveil the nature of the stellar
overdensities. In the present cases we use 2MASS $\jj\times\jh$ and $\jj\times\jk$ CMDs extracted
from a central region that, in the case of star clusters, should provide high-contrast CMDs. However,
features present in the central CMDs and in the respective comparison field, show that field stars
are an important component in the CMDs (Sect.~\ref{Disc}). Thus, it is essential to quantitatively
assess the relative densities of field stars and potential cluster sequences.

To objectively quantify the field-star contamination in the CMDs we apply the statistical algorithm described 
in \citet{BB07}. It measures the relative number-densities of probable field and cluster stars in cubic CMD
cells whose axes correspond to the magnitude \jj\ and the colours \jh\ and \jk. These are the 2MASS colours 
that provide the maximum variance among CMD sequences for OCs of different ages (e.g. \citealt{TheoretIsoc}).
The algorithm {\em (i)} divides the full range of CMD magnitude and colours into a 3D grid, {\em (ii)} computes 
the expected number-density of field stars in each cell based on the number of comparison field stars with
similar magnitude and colours as those in the cell, and {\em (iii)} subtracts the expected number of field
stars from each cell. By construction, the algorithm is sensitive to local variations of field-star 
contamination with colour and magnitude. Typical cell dimensions are $\Delta\jj=0.5$, and
$\Delta\jh=\Delta\jk=0.25$, which are large enough to allow sufficient star-count statistics in individual
cells and small enough to preserve the morphology of different CMD evolutionary sequences. As comparison
field we use the region $\rl<R<\rx$ around the cluster centre to obtain representative background star-count
statistics, where \rl\ is the limiting radius (Sect.~\ref{RDPs}). Further details on the algorithm, including
discussions on subtraction efficiency and limitations, are given in \citet{BB07}. The algorithm also gives
the parameter \ns\ which, for a given extraction, corresponds to the ratio of the number of stars in the
decontaminated CMD with respect to the $\rm1\sigma$ Poisson fluctuation measured in the observed CMD. CMDs 
of star clusters should have \ns\ significantly larger than 1. 

\begin{figure}
\resizebox{\hsize}{!}{\includegraphics{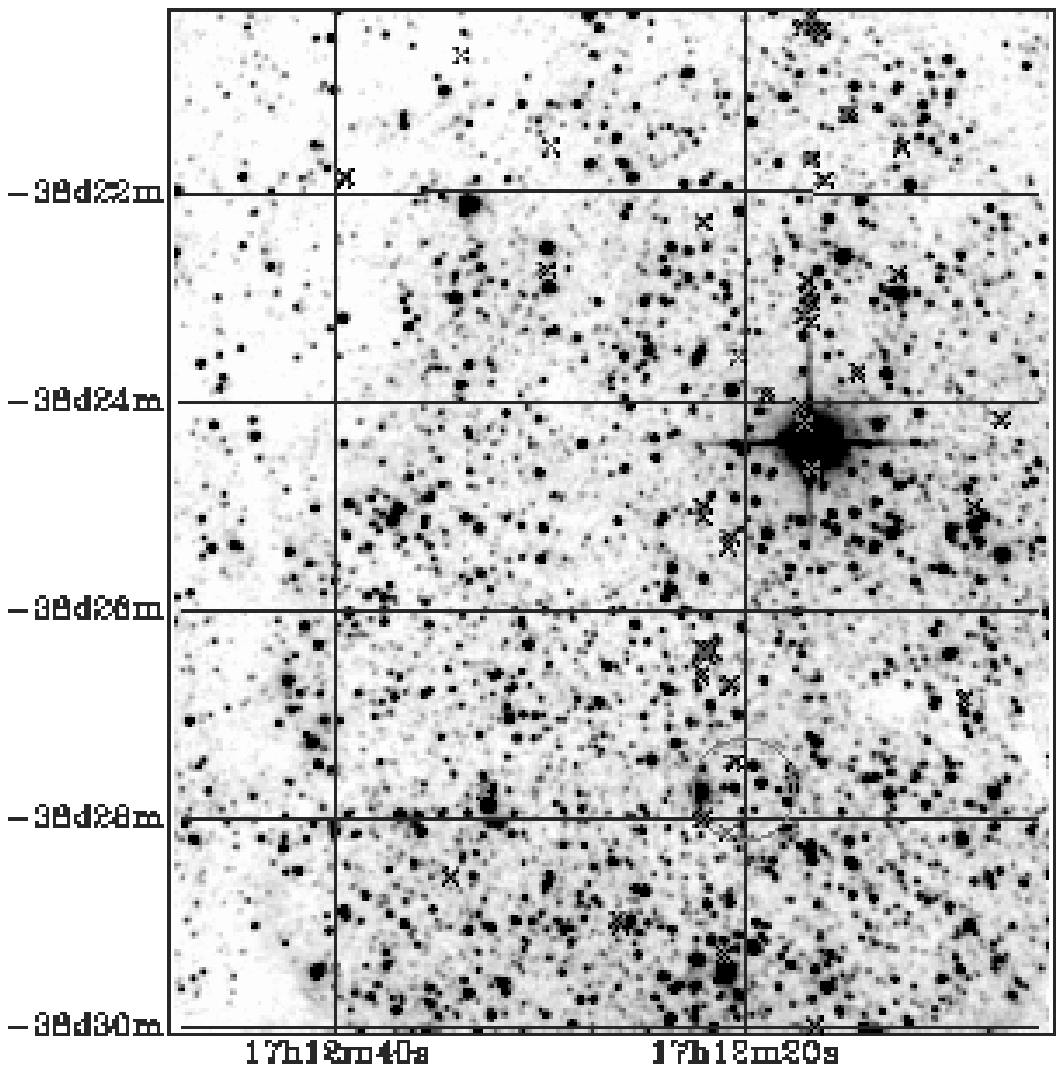}}
\caption[]{2MASS $10\arcmin\times10\arcmin$ \ks\ image of FSR\,1755.}
\label{fig2}
\end{figure}


\subsection{Fundamental parameters}
\label{FundPar}

Fundamental parameters are derived with solar-metallicity Padova isochrones (\citealt{Girardi02}) computed 
with the 2MASS \jj, \hh\ and \ks\ filters\footnote{\em stev.oapd.inaf.it/$\sim$lgirardi/cgi-bin/cmd }. The 
2MASS transmission filters produced isochrones very similar to the Johnson-Kron-Cousins (e.g. \citealt{BesBret88}) 
ones, with differences of at most 0.01 in \jh\ (\citealt{TheoretIsoc}).

The isochrone fit gives the age and the reddening \ejh, which converts to \ebv\ and \aV\ through the
transformations $A_J/A_V=0.276$, $A_H/A_V=0.176$, $A_{K_S}/A_V=0.118$, and $A_J=2.76\times\ejh$
(\citealt{DSB2002}), assuming a constant total-to-selective absorption ratio $R_V=3.1$. We also compute
the distance from the Sun (\ds) and the Galactocentric distance (\dgc), based on the recently derived value
of the Sun's distance to the Galactic centre $\Rgc=7.2$\,kpc (\citealt{GCProp}). Age, \aV, \ds\ and \dgc\ 
are given in cols.~5 to 8 of Table~\ref{tab2}, respectively.

\begin{table*}
\caption[]{Fundamental parameters from the present paper}
\label{tab2}
\renewcommand{\tabcolsep}{4.35mm}
\renewcommand{\arraystretch}{1.2}
\begin{tabular}{lccccccc}
\hline\hline
Target&$\alpha(2000)$&$\delta(2000)$&\ns&Age&\aV&\ds&\dgc\\
&(hms)&($^\circ\,\arcmin\,\arcsec$)&&(Gyr)&(mag)&(kpc)&(kpc)\\
(1)&(2)&(3)&(4)&(5)&(6)&(7)&(8)\\
\hline
Ru\,101, FSR\,1603&12:09:45&$-$62:59:17&$6.1\pm0.8$&$1.0\pm0.1$&$1.7\pm0.1$&$2.7\pm0.2$&$6.4\pm0.3$\\
FSR\,1755&($\ddagger$)& ($\ddagger$)&$5.8\pm0.9$&$<5\,Myr$&$\approx4.1$&$\approx1.4$&$\approx5.8$\\
\hline
\end{tabular}
\begin{list}{Table Notes.}
\item Cols.~2 and 3: Optimised central coordinates (Sect.~\ref{2mass}); $(\ddagger)$ indicates 
same central coordinates as in \citet{FSRcat}. Col.~4: Ratio of
the decontaminated star-counts to the $\rm1\sigma$ fluctuation level of the observed photometry.
Col.~6: $\rm\aV=3.1\,\ebv$. Col.~8: \dgc\ calculated using $\Rgc=7.2$\,kpc (\citealt{GCProp}) as the
distance of the Sun to the Galactic centre.
\end{list}
\end{table*}

\subsection{Colour-magnitude filters}
\label{CMF}

Colour-magnitude filters are used to exclude stars with colours compatible with those of the
foreground/background field. They are wide enough to accommodate cluster MS and evolved star
colour distributions, allowing for the $1\sigma$ photometric uncertainties. Colour-magnitude
filter widths should also account for formation or dynamical evolution-related effects, such
as enhanced fractions of binaries (and other multiple systems) towards the central parts of
clusters, since such systems tend to widen the MS (e.g. \citealt{BB07}; \citealt{N188};
\citealt{HT98}; \citealt{Kerber02}).

However, residual field stars with colours similar to those of the cluster are expected to remain 
inside the colour-magnitude filter region. They affect the intrinsic stellar radial distribution 
profile to an extent that depends on the relative densities of field and cluster stars. The 
contribution of the residual contamination to the observed RDP is statistically taken into account 
by means of the comparison field. 

\subsection{Stellar radial density profiles}
\label{RDPs}

Star clusters usually have RDPs that follow some well-defined analytical profile.
The most often used are the single-mass, modified isothermal sphere of \citet{King66}, the modified
isothermal sphere of \citet{Wilson75}, and the power-law with a core of \citet{EFF87}. Each function
is characterised by different parameters that are somehow related to cluster structure. However,
considering that error bars in the present RDPs are significant (see below), and that our goal here
is basically to determine the nature of the targets, we decided for the
analytical profile $\sigma(R)=\sigma_{bg}+\sigma_0/(1+(R/R_C)^2)$, where $\sigma_{bg}$ is
the residual background density, $\sigma_0$ is the central density of stars, and \rc\ is the core
radius. This function is similar to that introduced by \cite{King1962} to describe the surface
brightness profiles in the central parts of globular clusters.

The stellar RDPs are built with colour-magnitude filtered photometry (Sect.~\ref{CMF}). Minimisation 
of the presence of non-cluster stars by the colour-magnitude filter results in RDPs with a significantly
higher contrast with the background (e.g. \citealt{BB07}).
To avoid oversampling near the centre and undersampling at large radii, RDPs are built by counting
stars in rings of increasing width with distance to the centre. The number and width of the rings are
adjusted to produce RDPs with adequate spatial resolution and as small as possible $1\sigma$ Poisson
errors. The residual background level of each RDP corresponds to the average number of colour-magnitude
filtered stars measured in the comparison field. The $R$ coordinate (and respective uncertainty) of each
ring corresponds to the average position and standard deviation of the stars inside the ring.

We also compute the density contrast parameter $\delta_c=1+\sigma_0/\sigma_{bg}$, that measures
the intrinsic contrast of a cluster RDP (in the central region) with the background. In general
terms, $\delta_c$ is a measure of the detectability of clusters by means of star counts. Obviously,
the presence of O stars in very young clusters may improve the detectability, even for low-$\delta_c$
cases. As a caveat we note that $\delta_c$ values measured in colour-magnitude filtered RDPs do not
necessarily correspond to the visual contrast produced by optical/IR images. The minimisation of
field star contamination produces RDPs with higher contrast than those resulting from the raw
photometry (e.g. \citealt{FaintOCs}).

As a measure of the cluster size, we also derive the limiting radius and uncertainty, which are
estimated by comparing the RDP (taking into account fluctuations) with the background level. \rl\
corresponds to the distance from the cluster centre where RDP and background become statistically
indistinguishable. For practical purposes, most of the cluster stars are contained within $\rl$
(Fig.~\ref{fig4}). The limiting radius should not be mistaken for the tidal radius; the latter
is usually derived from King (or other analytical functions) fits to RDPs, which depend on wide
surrounding fields and as small as possible Poisson errors (e.g. \citealt{BB07}). In contrast,
\rl\ comes from a visual comparison of the RDP and background levels.

The empirical determination of a cluster-limiting radius depends on the relative levels of RDP and
background (and respective fluctuations). Thus, dynamical evolution may indirectly affect the
measurement of the limiting radius. Since mass segregation preferentially drives low-mass stars to the
outer parts of clusters, the cluster/background contrast in these regions tends to lower as clusters age.
As an observational consequence, smaller values of limiting radii should be measured, especially for
clusters in dense fields. However, simulations of King-like OCs (\citealt{BB07}) show that, provided not
exceedingly high, background levels may produce limiting radii underestimated by about 10--20\%. The core
radius, on the other hand, is almost insensitive to background levels (\citealt{BB07}). This occurs because
\rc\ results from fitting the King-like profile to a distribution of RDP points, which minimises background
effects.

\section{Discussion}
\label{Disc}

\subsection{The IAC nature of FSR\,1603}
\label{FSR1603}

\begin{figure}
\resizebox{\hsize}{!}{\includegraphics{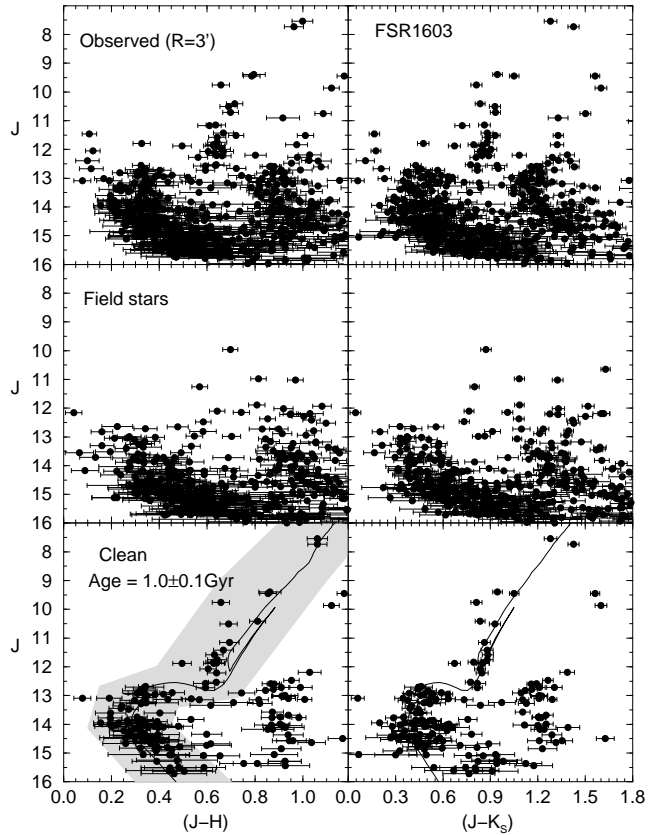}}
\caption{2MASS CMDs extracted from the $R<3\arcmin$ region of FSR\,1603. Top panels: observed photometry
with the colours $\jj\times\jh$ (left) and $\jj\times\jk$ (right). Middle: equal-area comparison field.
Besides some contamination of disk and bulge stars, a populous MS and a conspicuous cluster red giant 
clump show up in the upper panels. Bottom panels: decontaminated CMDs set with the 1\,Gyr Padova isochrone 
(solid line). The colour-magnitude filter used to isolate cluster MS/evolved stars is shown as a shaded 
region.}
\label{fig3}
\end{figure}

\begin{figure}
\resizebox{\hsize}{!}{\includegraphics{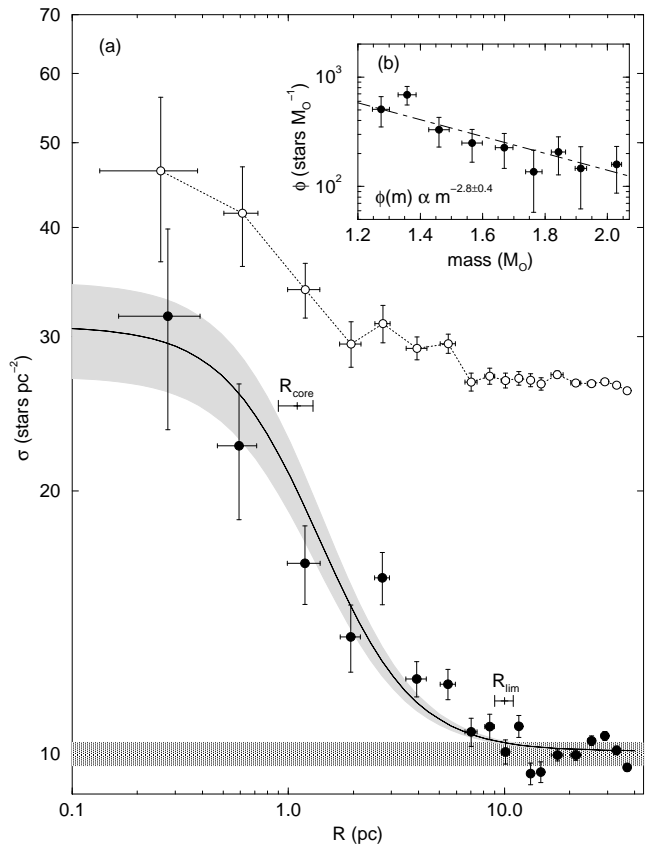}}
\caption{Panel (a): Stellar RDP (filled circles) of FSR\,1603 built with colour-magnitude photometry.
Solid line: best-fit King profile. Horizontal shaded region: offset field stellar background level. 
Gray regions: $1\sigma$ King fit uncertainty. The core and limiting radii are shown. The RDP built 
with the observed photometry (open circles)  is shown for comparison. Absolute scale is used. Panel 
(b): Field-star decontaminated mass function for the MS stars, fitted with $\phi(m)\propto m^{-2.8}$.}
\label{fig4}
\end{figure}

CMDs of the central ($R<3\arcmin\sim2\times\rc$) region of FSR\,1603 are presented in Fig.~\ref{fig3}
(top panels). This extraction contains the bulk of the cluster stars (Fig.~\ref{fig4}). Features that 
stand out are a cluster-like population corresponding to a populous MS ($0.2\la\jh\la0.6$ and 
$12.5\la\jj\la16$), a giant clump ($0.6\la\jh\la0.7$ and $12.2\la\jj\la11$) and a giant branch 
($0.7\la\jh\la1.0$ and $\jj\la11$), together with a redder sequence ($\jh\ga0.8$). Some of these 
features (especially around the MS and the red component) are present as well in the equal-area 
comparison field (middle panels), extracted from a ring near $30\arcmin$, which suggests some contamination 
by disk and bulge stars. However, the central and comparison field extractions present significant 
differences in terms of CMD densities. Besides, the giant clump and giant branch are not present in 
the comparison field. Similar features are present in the $\jj\times\jk$ CMDs (right panels). 
The qualitative discussion above strongly suggests that FSR\,1603 is an intermediate-age OC.

The resulting field star decontaminated CMDs of FSR\,1603 are shown in the bottom panels of Fig.~\ref{fig3}. 
The comparison field was taken from the region $20\arcmin\leq R\leq 50\arcmin$. Bulge and disk contamination 
have been properly taken into account,  except for about 40 red stars that remain in the decontaminated 
CMDs. Considering that the original red component contains about 214 stars (top panels), the statistical
significance of the leftovers represent $\ns=2.7$, significantly lower than that of stellar sequence
identified with FSR\,1603 (Table~\ref{tab1}). As shown in \citet{ProbFSR}, such low values of \ns\ are 
typical of field fluctuations detected as overdensities by \citet{FSRcat}. What emerges from the 
$\ns=6.1\pm0.8$ decontaminated CMDs are conspicuous sequences, especially the giant clump and the MS, 
typical of a relatively populous Gyr-class OC.

Indeed, the decontaminated CMD morphology of FSR\,1603 is well represented by the 1\,Gyr isochrone, 
with a distance modulus $\mMJ=12.6\pm0.1$, and a reddening $\ejh=0.17\pm0.01$, which corresponds to 
$\ebv=0.54\pm0.04$ and $\aV=1.69\pm0.12$. Such values imply a distance from the Sun $\ds=2.7\pm0.1$\,kpc
and a Galactocentric distance $\dgc=6.4\pm0.2$\,kpc. At this distance, the scale is $\rm1\arcmin=0.774$\,pc.
The isochrone fit to FSR\,1603 is shown in the bottom panels of Fig.~\ref{fig3}.

Figure~\ref{fig4} shows the stellar RDP of FSR\,1603. Besides the RDP resulting from the colour-magnitude 
filter, we also show, for illustrative purposes, that produced with the observed (raw) photometry. In both
cases, the central region presents a significant density contrast with respect to the background, especially
the colour-magnitude filter RDP, with $\delta_c=3.0\pm0.4$. The adopted King-like function describes well
the latter RDP throughout the full radii range, within uncertainties. We derive a core radius
$\rm\rc=1.1\pm0.2\,pc=1.4\arcmin\pm0.3\arcmin$ and a limiting radius $\rm\rl=10\pm1\,pc=13\arcmin\pm1.3\arcmin$. 
The central density of stars $\rm\sigma_0=20\pm4\,stars\,pc^{-2}$ and the core radius (\rc) are derived from 
the RDP fit, while the background level $\rm\sigma_{bg}=10\pm0.3\,stars\,pc^{-2}$ is measured in the respective 
comparison field. The best-fit solution is superimposed on the colour-magnitude filtered RDP (Fig.~\ref{fig4}). 
Because of the 2MASS photometric limit, which for FSR\,1603 corresponds to a cutoff for stars brighter than
$\jj\approx16$ (Fig.~\ref{fig3}), $\sigma_0$ should be taken as a lower limit to the actual central number-density.
Within the uncertainty, the present value of \rc\ agrees with that given by \citet{FSRcat}.

The relatively populous nature of the MS of FSR\,1603 can be used to estimate the mass stored
in stars. Considering the field-star decontaminated photometry, the number of observed MS and
evolved stars for $\rm R<\rl$ amount to $\rm N^{MS}_{obs}=279\pm53$ and 
$\rm N^{Evol}_{obs}=46\pm10$, respectively. Using the mass-luminosity relation derived from the
1\,Gyr Padova isochrone set with the parameters presented above, the observed MS mass is
$\rm M^{MS}_{obs}=407\pm46\,\ms$. For the evolved stars we multiply $\rm N^{Evol}_{obs}$ by
the stellar mass at the TO ($\rm m_{TO}\approx2.1\,\ms$), which yields the estimate
$\rm M^{Evol}_{obs}=96\pm20\,\ms$. Thus, the total mass stored in the observed stars of FSR\,1603
amounts to $\rm M_{obs}=503\pm50\,\ms$.

We also build the mass function (MF) $\phi(m)=\frac{dN}{dm}$, for the MS stars
following the methods discussed in \citet{DetAnalOCs}. To maximise the statistical significance
of field-star counts we take as the comparison field the region at $\rm 15\leq R(pc)\leq 39$,
that lies $\rm\geq5\,pc$ beyond the limiting radius. First we apply the colour-magnitude filter
(Fig.~\ref{fig3}) to the stars located inside the limiting radius, which eliminates most of the
field stars, leaving a residual contamination. We deal with this statistically by building the
luminosity functions (LFs) for the cluster and comparison field separately.  \jj, \hh\ and \ks\
LFs are built by counting stars in magnitude bins from the respective faint magnitude limit to
the TO. To avoid undersampling near the turn-off and oversampling at the faint limit, magnitude
bins are wider in the upper MS than in the lower MS. Correction is made for the different solid
angles of the comparison field and cluster region. The intrinsic cluster LF is obtained by
subtracting the offset-field LF. Finally, the intrinsic LF is transformed into MF using the
mass-luminosity relation obtained from the 1\,Gyr Padova isochrone and the observed distance
modulus $\mMJ=12.6$. These procedures are applied independently to the three 2MASS bands. The
final MF is produced by combining the \jj, \hh\ and \ks\ MFs. Further details on MF construction
are given in \citet{FaintOCs}. The resulting MF, covering the mass range
$\rm 1.27\,\ms\leq m_{MS}\leq2.03\,\ms$, is shown in the inset of Fig.~\ref{fig4}, where we also
include the fit with the function $\phi(m)\propto m^{-(1+\chi)}$. Within the uncertainty, the 
slope $\chi=1.8\pm0.4$ is similar to the $\chi=1.35$ of the Initial Mass Function (IMF) of 
\citet{Salp55}.

Finally, with the above MF we can estimate the total stellar mass in FSR\,1603. For masses below the 
observed lower limit of 1.27\,\ms, we extrapolate the observed MF down to the H-burning mass limit 
($0.08\,\ms$) using the universal IMF of \citet{Kroupa2001}. The latter IMF 
assumes the slopes $\chi=0.3\pm0.5$ for the range $0.08\leq m(\ms)\leq0.5$ and $\chi=1.3\pm0.3$ for
$0.5\leq m(\ms)\leq1.0$. With such considerations, the total (extrapolated MS $+$ evolved) mass
amounts to $\rm M_{tot}=(2.3\pm0.9)\times10^3\,\ms$, which implies the average mass density
$\rm\rho=0.55\pm0.22\,\ms\,pc^{-3}$ and the projected density $\rm\sigma=7.3\pm2.9\,\ms\,pc^{-2}$.

As a caveat we note that the total mass estimates should be taken as upper limits, since because
of dynamical evolution, significant fractions of the low-mass content may have been lost to the
field (e.g. \citealt{OldOCs}).

\subsection{The possible embedded cluster FSR\,1755}
\label{FSR1755}

The images of FSR\,1755 (Figs.~\ref{fig1} and \ref{fig2}) clearly indicate that this object
is a complicated case. Indeed, the differential absorption due to the dust clouds produces
conspicuous variations in the stellar number-density, as can be seen in Fig.~\ref{fig6}. 


According to SIMBAD\footnote{http://simbad.u-starsbg.fr/simbad/ }, the bright star at
$\approx2\arcmin$ south of the central coordinates of FSR\,1755 is the O star
$CD-38^\circ\ 11636$.

FSR\,1755 appears to be spatially associated with the H\,II region Sh\,2-3 (\citealt{Sharp59}), 
located at
$\rm\alpha(J2000)=17:12:21$ and $\rm\delta(J2000)=-38:27:00$, also known as
Gum\,58 (\citealt{Gum55}) and RCW\,120 (\citealt{RCW60}). \citet{Russ03} provides the radial
velocity $\rm v_r=12.0\,km\,s^{-1}$, and the kinematic distance $\ds=1.8\pm0.7$\,kpc for
Sh\,2-3. According to the spatial distribution of the spiral arms (\citealt{Russ03}), Sh\,2-3
appears to be related to the Sagittarius-Carina arm. Sh\,2-3 is an optical H\,II region of dimensions
$\sim\rm 7\arcmin\ (EW) \times\ 10\arcmin\ (NS)$, excited by the O\,8V star  $CD-38^\circ\ 11636$
(LSS\,3959), at $\rm\alpha(J2000)=17:12:20.6$ and $\rm\delta(J2000)=-38:29:26$. It was
identified spectroscopically by \citet{GG70}. \citet{VM75} provide $\ds=1.7$\,kpc as the
photometric distance to this star. $CD-38^\circ\ 11636$ has the apparent magnitudes
$B=11.93$ and $V=10.79$ (\citealt{Zavagno07}), $\jj=8.013\pm0.021$, $\hh=7.708\pm0.036$,
and $\ks=7.523\pm0.020$ (2MASS). Based on the optical photometry of $CD-38^\circ\ 11636$, \citet{Zavagno07}
compute $\aV=4.36$ and estimate the distance to this star as $\ds=1.34$\,kpc, which agrees
with their kinematic distance of $\ds=1.35$\,kpc. With the 2MASS photometry of $CD-38^\circ\ 11636$
and the absolute \jj\ and \hh\ magnitudes of an O\,8V synthetic star (\citealt{MarPle06})
we compute $\ds=1.35$\,kpc, which agrees with the previous values. We also derive $\ejh=0.4$,
$\rm A_J=1.14$, $\aV=4.15$, and $\mMJ=11.9$.

\begin{figure}
\resizebox{\hsize}{!}{\includegraphics{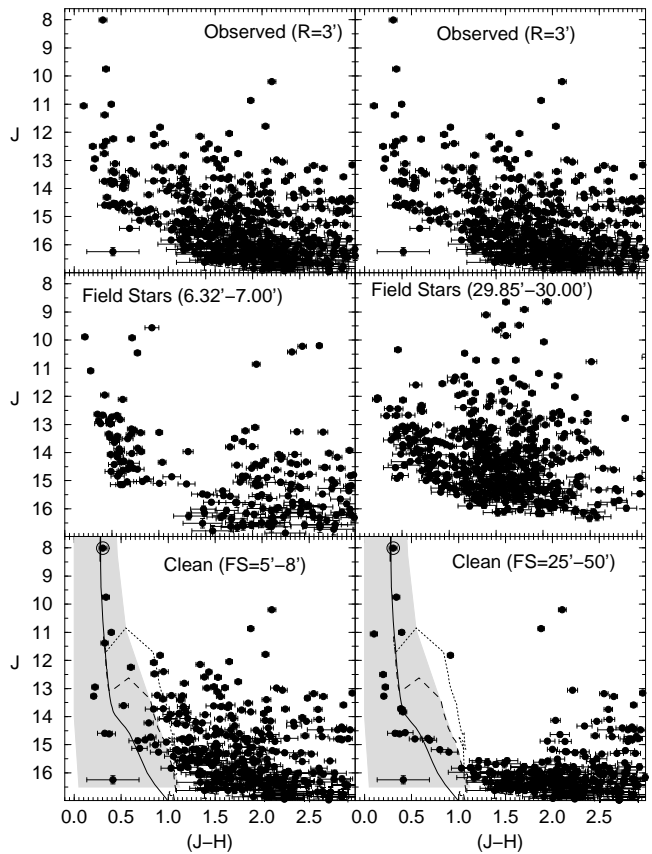}}
\caption{Top panels: $\jj\times\jh$ CMDs of the central $\rm R=3\arcmin$ region of FSR\,1755.
Middle panels: Equal-area comparison fields extracted from 2 different regions. Bottom panels: 
CMDs decontaminated using the different comparison fields. The O\,8V star $CD-38^\circ\ 11636$
is shown by the circle. The 3\,Myr Padova isochrone (heavy solid line), together with the 1\,Myr 
(dotted) and 5\,Myr (dashed) PMS isochrones have been set for $\ds=1.35$\,kpc and $\ebv=1.3$. 
The shaded region represents the colour-magnitude filter.}
\label{fig5}
\end{figure}

In Fig.~\ref{fig5} (top panels) we show the $\jj\times\jh$ CMD extracted from the $R=3\arcmin$
central region of FSR\,1755. An almost vertical, blue ($0.05\la\jh\la0.5$, $8\la\jj\la15$) 
sequence --- typical of very young populations --- appears to merge into a dominant concentration 
of red stars. The strong differential reddening in the field of FSR\,1755 hinders the quantitative 
use of the decontamination algorithm. However, it can be applied as a qualitative probe on CMD 
features, for different choices of the comparison field. Two cases are examined, one taking field 
stars from the region $5\arcmin\leq R\leq 8\arcmin$, that basically corresponds to the deep RDP 
depression (Fig.~\ref{fig6}), and the other from an external region $25\arcmin\leq R\leq 50\arcmin$ away 
from the centre.  For visual purposes only, we characterise the field contamination on the  
$R=3\arcmin$ central CMDs, resulting from both choices of comparison field, by means of the equal-area 
extractions $6.32\arcmin\leq R\leq 7\arcmin$ and $29.85\arcmin\leq R\leq 30\arcmin$, respectively. 
They are shown in the middle panels of Fig.~\ref{fig5}. For statistical representativity of the field 
stars, the decontamination algorithm uses the full radial scale of the adopted comparison fields, 
$5\arcmin\leq R\leq 8\arcmin$ and $25\arcmin\leq R\leq 50\arcmin$, respectively.  In both cases, 
the stellar distributions are similar in terms of colour and magnitudes, but they differ significantly 
in density. 

\begin{figure}
\resizebox{\hsize}{!}{\includegraphics{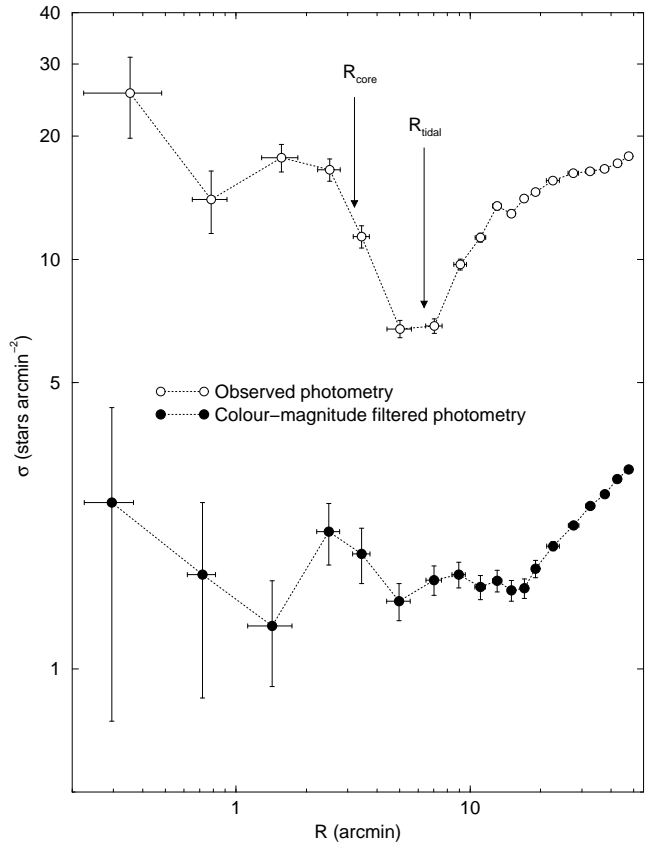}}
\caption{Observed (open circles) and colour-magnitude filtered (filled circles) RDPs of 
FSR\,1755. The core and tidal radii computed by \citet{FSRcat} are indicated. The 
absorption-related depression at $R\approx5\arcmin - 7\arcmin$ helps create the impression 
of a GC.}
\label{fig6}
\end{figure}

Consequently, the resulting decontaminated CMDs also present differences, especially in the
remaining field-star distribution. However, the blue sequence consistently remains in both
CMDs. 

Assuming that FSR\,1755 is associated with Sh\,2-3, we apply to the decontaminated CMDs the
3\,Myr Padova isochrone as representative of the young age, together with the pre-MS (PMS) 
tracks of \citet{Siess2000} with the ages 1\,Myr and 5\,Myr, set for $\ds=1.35$\,kpc and 
$\ebv=1.3$. What results in both cases is consistent with a sparse embedded cluster. Thus, 
most of what remains in the CMDs as the red component results from the strong differential 
reddening in the rich field. However, because of the strong differential absorption, we cannot 
exclude the possibility that some fraction of the red component is composed of very reddened 
PMS stars.

Finally, in Fig.~\ref{fig6} we present the RDPs of FSR\,1755. The deep RDP depression, which 
is due to dust absorption, associated with the large number of stars related to the red component,
creates the impression of a high central stellar concentration and large core radius. Taken together,
these features match the GC criterion established by \citet{FSRcat}. Indeed, the RDP dip (Fig.~\ref{fig6})
coincides with the tidal radius computed by \citet{FSRcat}.

On the other hand, most of the above features are absent in the colour-magnitude filtered RDP, which 
has a low-level central concentration of stars typical of a poorly-populated star cluster, and a small 
bump at $R\approx2.5\arcmin$ that coincides with the radial distance of $CD-38^\circ\ 11636$.

From the above we conclude that the potentially populous cluster FSR\,1755 is in fact an artifact 
produced by the highly spatially variable star-count density and absorption across the field. However,
we cannot exclude the possibility of a sparse young cluster embedded in the H\,II region Sh\,2-3. 

\section{Summary and conclusions}
\label{Conclu}

In this paper we investigated the nature of the GC candidates FSR\,1603 and FSR\,1755, taken from
the \citet{FSRcat} catalogue, by means of decontaminated \jj\, \hh\ and \ks\ 2MASS CMDs and
stellar RDPs. Both objects were originally classified as GC candidates by \citet{FSRcat} based
essentially on the number of cluster stars, the core radius and the central star density.
Basically, what results from the present work is that neither object is in fact a globular 
cluster: FSR\,1603 is a populous intermediate-age open cluster and FSR\,1755 may consist of a 
sparse young cluster embedded in a complex H II region. 

FSR\,1603 corresponds to the OC Ruprecht\,101, for which cluster parameters were derived for the 
first time. It is a massive IAC, with $\rm age=1.0\pm0.1$\,Gyr, located at $\ds=2.7\pm0.2$\,kpc 
from the Sun and Galactocentric distance $\dgc=6.3\pm0.3$\,kpc, reddening $\ebv=0.54\pm0.04$, core 
and limiting radii $\rc=1.1\pm0.2$\,pc and $\rl=10\pm1$\,pc, respectively. The mass stored in the 
observed stars is $\rm M_{obs}\approx500\,\ms$; extrapolation to the low-MS stars ($\sim0.08\,\ms$) 
increases the mass to $\rm M_{tot}\sim2\,300\,\ms$. Ru\,101 was probably misclassified as a GC 
candidate because it is a populous OC.

FSR\,1755 is related to the optical H\,II region Sh\,2-3, located at $\ds\approx1.4$\,kpc. 
The object is surrounded both by $\rm H_\alpha$ emission and dust absorption even in the \ks\ 
band. A nearly vertical, blue CMD sequence, that includes the O\,8V star $CD-38^\circ\ 11636$, 
appears to define FSR\,1755 as a sparse embedded
cluster. Because of the strong differential reddening in the surroundings, a large number of
red stars show up in the CMDs, which probably produced the stellar enhancement that was taken 
as GC candidate by \citet{FSRcat}. We conclude that the GC identification of FSR\,1755 is a 
dust-absorption artifact in a very rich stellar field.

Systematic surveys such as that of \citet{FSRcat} are important to detect new star cluster
candidates throughout the Galaxy. Nevertheless, works like the present one, that fine tune
the analysis by means of field-star decontaminated CMDs and stellar radial profiles, turn
out to be fundamental to probe the nature of such candidates, especially those projected 
against dense stellar fields.

\section*{acknowledgements}
We thank the anonymous referee for suggestions.
We acknowledge partial support from CNPq (Brazil). This research has made use of the SIMBAD database,
operated at CDS, Strasbourg, France, as well as of data products from the Two Micron All Sky Survey
(2MASS), which is a joint project of the University of Massachusetts and the Infrared Processing and
Analysis Center/California Institute of Technology, funded by the National Aeronautics and Space
Administration and the National Science Foundation. We employed Digitized Sky Survey images from the
Space Telescope Science Institute (U.S. Government grant NAG W-2166) obtained using the extraction
tool from CADC (Canada). We also made use of the WEBDA database, operated at the Institute for Astronomy
of the University of Vienna.



\begin{thebibliography}{}

\bibitem[\protect\citeauthoryear{Bessell \& Brett}{1988}]{BesBret88}
   Bessell, M.S. \& Brett, J.M. 1988, PASP, 100, 1134
   
\bibitem[\protect\citeauthoryear{Bica et al.}{2006a}]{FaintOCs}
   Bica, E., Bonatto, C. \& Blumberg, R. 2006a, A\&A, 460, 83

\bibitem[\protect\citeauthoryear{Bica et al.}{2006}]{GCProp}
   Bica, E., Bonatto, C., Barbuy, B. \& Ortolani, S. 2006, A\&A, 450, 105
   
\bibitem[\protect\citeauthoryear{Bica et al.}{2007}]{FSR584}
   Bica, E., Bonatto, C., Ortolani, S. \& Barbuy, B. 2007, A\&A, 472, 483

\bibitem[\protect\citeauthoryear{Bica, Bonatto \& Camargo}{2007}]{ProbFSR}
   Bica, E., Bonatto, C. \& Camargo, D.. 2007, MNRAS, in press, (astro-ph/0712.0762)

\bibitem[\protect\citeauthoryear{Binney \& Merrifield}{1998}]{Binney1998}
   Binney, J. \& Merrifield, M. 1998, in {\em Galactic Astronomy}, Princeton,
   NJ: Princeton University Press. (Princeton series in Astrophysics)

\bibitem[\protect\citeauthoryear{Bonatto, Bica \& Girardi}{2004}]{TheoretIsoc}
   Bonatto, C., Bica, E. \& Girardi, L. 2004, A\&A, 415, 571
   
\bibitem[\protect\citeauthoryear{Bonatto, Bica \& Santos Jr.}{2005}]{N188}
   Bonatto, C., Bica, E. \&  Santos Jr., J.F.C. 2005, A\&A, 433, 917

\bibitem[\protect\citeauthoryear{Bonatto \& Bica}{2005}]{DetAnalOCs}
    Bonatto, C. \&  Bica, E. 2005, A\&A, 437, 483

\bibitem[\protect\citeauthoryear{Bonatto \& Bica}{2007a}]{OldOCs}
   Bonatto, C. \& Bica, E. 2007a, A\&A, 473, 445

\bibitem[\protect\citeauthoryear{Bonatto \& Bica}{2007b}]{BB07}
   Bonatto, C. \& Bica, E. 2007b, MNRAS, 377, 1301

\bibitem[\protect\citeauthoryear{Bonatto \& Bica}{2007c}]{Pap11GCs}
   Bonatto, C. \& Bica, E. 2007c, A\&A, accepted, (astro-ph:0711.1434)
   
\bibitem[\protect\citeauthoryear{Bonatto et al.}{2007}]{FSR1767}
   Bonatto, C., Bica, E., Ortolani, S. \& Barbuy, B. 2007, MNRAS, 381, L45
   
\bibitem[\protect\citeauthoryear{Dutra, Santiago \& Bica}{2002}]{DSB2002}
   Dutra, C.M., Santiago, B.X. \& Bica, E. 2002, A\&A, 383, 219

\bibitem[\protect\citeauthoryear{Elson, Fall \& Freeman}{1987}]{EFF87}
   Elson, R.A.W., Fall, S.M. \& Freeman, K.C. 1987, ApJ, 323, 54

\bibitem[\protect\citeauthoryear{Froebrich, Meusinger \& Scholz}{2007}]{FSR1735}
   Froebrich, D., Meusinger, H. \& Scholz, A. 2007, MNRAS, 377, L54

\bibitem[\protect\citeauthoryear{Froebrich, Scholz \& Raftery}{2007}]{FSRcat}
   Froebrich, D., Scholz, A. \& Raftery, C.L. 2007, MNRAS, 374, 399
   
\bibitem[\protect\citeauthoryear{Froebrich, Meusinger \& Davis}{2007}]{FSR190}
   Froebrich, D., Meusinger, H. \& Davis, C.J. 2007, MNRAS, in press 
   (astro-ph:0710-2030)

\bibitem[\protect\citeauthoryear{Georgelin \& Georgelin}{1970}]{GG70}
   Georgelin, Y.P. \& Georgelin, Y.M. 1970, A\&ASS, 3, 1

\bibitem[\protect\citeauthoryear{Girardi et al.}{2002}]{Girardi02}
   Girardi, L., Bertelli, G., Bressan, A., et al. 2002, A\&A, 391, 195

\bibitem[\protect\citeauthoryear{Gum}{1955}]{Gum55}
   Gum, C.S. 1955, MmRAS, 67, 155

\bibitem[\protect\citeauthoryear{Hurley \& Tout}{1998}]{HT98}
   Hurley, J. \& Tout, A.A. 1998, MNRAS, 300, 977

\bibitem[\protect\citeauthoryear{Kerber et al.}{2002}]{Kerber02}
   Kerber, L.O., Santiago, B.X., Castro, R. \& Valls-Gabaud, D. 2002, A\&A, 390, 121

\bibitem[\protect\citeauthoryear{King}{1962}]{King1962}
   King, I. 1962, AJ, 67, 471

\bibitem[\protect\citeauthoryear{King}{1966}]{King66}
   King, I. 1966, AJ, 71, 64
   
\bibitem[\protect\citeauthoryear{Kroupa}{2001}]{Kroupa2001}
   Kroupa, P. 2001, MNRAS, 322, 231

\bibitem[\protect\citeauthoryear{Martins \& Plez}{2006}]{MarPle06}
   Martins, F. \& Plez, B. 2006, A\&A, 457, 637

\bibitem[\protect\citeauthoryear{Mermilliod}{1996}]{Merm1996}
   Mermilliod, J.C. 1996, in {\it The Origins, Evolution, and Destinies of
   Binary Stars in Clusters}, ASP Conference Series, eds. E.F. Milone \& J.-C.
   Mermilliod, 90, 475

\bibitem[\protect\citeauthoryear{Rodgers, Campbell \& Whiteoak}{1960}]{RCW60}
   Rodgers, A.W., Campbell, C.T. \& Whiteoak, J.B. 1960, MNRAS, 121, 103

\bibitem[\protect\citeauthoryear{Russeil}{2003}]{Russ03}
   Russeil, D. 2003, A\&A, 397, 133

\bibitem[\protect\citeauthoryear{Salpeter}{1955}]{Salp55}
   Salpeter, E. 1955, ApJ, 121, 161
   
\bibitem[\protect\citeauthoryear{Sharpless}{1959}]{Sharp59}
   Sharpless, S. 1959, ApJS, 4, 257
   
\bibitem[\protect\citeauthoryear{Siess, Dufour \& Forestini}{2000}]{Siess2000}
   Siess, L., Dufour, E. \& Forestini, M. 2000, A\&A, 358, 593
   
\bibitem[\protect\citeauthoryear{Vogt \& Moffat}{1975}]{VM75}
   Vogt, N. \& Moffat, A.F.J. 1975, A\&A, 45, 405 

\bibitem[\protect\citeauthoryear{Wilson}{1975}]{Wilson75}
   Wilson, C.P. 1975, AJ, 80, 175

\bibitem[\protect\citeauthoryear{Zavagno et al.}{2007}]{Zavagno07}
   Zavagno, A., Pomar\`es, M., Deharveng, L., Hosokawa, Y., Russeil, D.
   \& Caplan, J. 2007, A\&A, 472, 835

\end{thebibliography}
\end{document}